\documentclass{an}
\usepackage{graphicx}
\usepackage{times}
\usepackage{fancyhdr}
\usepackage{natbib}

\begin{document}

\title{The statistics of Sco X-1 kHZ QPOs}

\author{Tomasz Bulik et al.}
\institute{Nicolaus Copernicus Astronomical Center, 
Bartycka 18, 00716 Warszawa, Poland}

\date{Received; accepted; published online}

\abstract{Recently  an additional technique was
applied to investigate the properties of kHz QPOS, i.e.
the analysis of the distribution of frequency ratios or 
frequencies themselves. 
I review the results of such work
on the data  from  Sco X-1: \citet{2003A&A...404L..21A}, which was later
criticized by\citep{2005A&A...437..209B}.
I find that the  findings of the  latter paper 
are consistent with 
results presented earlier: kHz QPOs cluster 
around the value corresponding to the frequency ratio of 2/3.
 I also discuss the random walk model of kHz 
 QPOs and possible future observations needed to verify it.
\keywords{}}

\correspondence{bulik@camk.edu.pl}

\maketitle

\section{Introduction}

The nature of kHz quasi periodic oscillations 
continues to puzzle astrophysicists. 
In order to unveil the nature 
of these phenomena different quatities 
are considered. For example the difference
of the two kHz QPO, which in the beat frequency model
is constant was analyzed.
The disk resonance model e.g. \citet{2001A&A...374L..19A}
predicts that the ratio of the two
kHz QPO frequencies  should assume a set of discete
values.
In order to test it  we  analyzed the statistical 
properties of KHz QPOs in Sco X-1 \citep{2003A&A...404L..21A}.
We have found that the frequency ratio
of the lower to higher QPO cluster around 
$0.688$, a value close to the canonical 
$2/3$ seen in the QPOs from  black hole candidates.
This discovery has recently been disputed by 
\citet{2005A&A...437..209B}. 

In this paper I will discuss the similarities 
and differences of the two analyses. 
In section 2 I present an overview of different 
methods used to compare distributions and in section 3
I summarize 
the results of \citet{2003A&A...404L..21A}.
I compare our results with the work
of \citep{2005A&A...437..209B} in section 4.
Section 5 contains summary and 
a discussion of the possible future 
observational tests.

\section{Comparison of distributions }

Inferring the shape of an underlying distribution 
given a set of observed quantities is a diffcult
task. In a number of cases we have no prior knowledge of what the shape of the distribution should be. Very often we do not know 
if the measured quantity is fundamental from the physical 
point of view. However, one can start by making 
simple assumptions about the data and testing them.

There is a number of ways to compare the shape of 
two distributions e.g. a model and an observed one.
The $chi^2$ test comparing a binned observed distribution with
a model are not very useful. Such comparison suffer from 
a number of arbitrary assumptions about the widths
of the bins which makes the results not reliable.
A different set of statistical tests that do not require
binning of the data is based on comparing cumulative distributions.
One can introduce various metrics to measures   distance 
between two cumulative distributions and calculate 
the probability distribution of such distance. 
A  very useful test using this approach is the Kolmogorov-Smirnov
test in which the metric  is defined as 
\begin{equation}
D_{KS}={\rm max} |(C_1 - C_2)|,
\end{equation}
where $C_i$ are the cumulative distributions.
The sensitivity of such test is limited, yet it serves as 
a convenient tool for comparing distributions and assessing
goodness of fit.
There exist modified and more sensitive versions of this 
test. For example the Fiszer-von Mises tests uses 
the metric defines as 
\begin{equation}
D_{Fisz}=\int (C_1 - C_2)^2,
\end{equation}
and is therefore much more sensitive to 
differences in overall shape of two distributions.
However calculating the probabilty distribution 
of $D_{Fisz}$ is a much more demanding task/

Any such test can be used for parameter 
estimation. Suppose that the model distribution 
$C_m(p_k)$
is a function of several parameters $p_k$.
We can then maximize the probability of the 
model given the observational set of data.
In this case however the most sensitive method 
to use is the maximum likelihood. 
In this approach we can calculate the probability
of the model $M$ given the data $X:(x_l)$:
\begin{equation}
{\cal L}(M|X) \propto \Pi {dp\over dx}(x_l).
\end{equation}
This likelihood function can be maximized with respect to the model
parameters to yield the best fit model.
The goodness of fit of such model can 
evaluated using e.g. the  Kolmogorov Smirnov test.
In order to find the   probability distribution
of the parameters one ususally assumes that 
it is proportional to
the likelihood function.

\section{Summary of the Sco X-1 analysis }

\begin{figure}
\resizebox{\hsize}{!}
{\includegraphics[width=0.9\columnwidth]{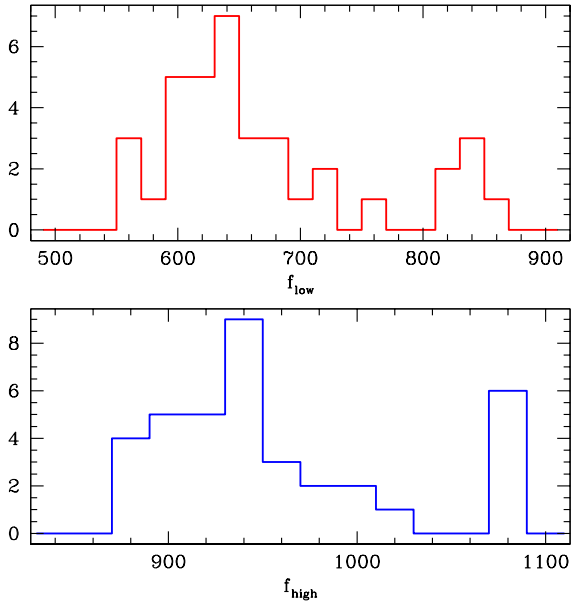}}
\caption{The binned distributions of the lower and upper QPO
frequencies from the data set of\citet{1997ApJ...481L..97V}. Note that the two distributions are similar but not identical.}
\label{fdist}
\end{figure}

In our paper we have addressed the question 
whether there is any evidence of clustering 
of the frequency ratios of kHz QPOs in Sco X-1. To do that 
we used a set of data published by \citet{1997ApJ...481L..97V}.
This set contained 39 frequency pairs and was used 
by the authors to prove that the difference
between the two frequencies is not constant.
We have first verified the hypothesis that
the distribution of frequency ratios is uniform using
the Kolmogorov-Smirnov test. This probability is 
\begin{equation}
P_{\rm uniform}=0.00047,
\end{equation}
which led us to consider the question if the frequency ratios
cluster. We have no a priori knowledge of the shape of the
distribution we are looking for as there is no
physical model that we could refer to. Therefore we decided
to model the shape of the distribution with 
a Lorentzian profile normalized on the interval $(0,1)$.
We have used the maximum likelihood method to find the best
fit which was a Lorentzian centered on $r_0=0.688$ with a width of
$\lambda=0.0151$. The Kolmogorov-Smirnov test 
provided a confirmation that the data could have been drawn 
from such distribution, since the probability was
\begin{equation}
P_{Lorentzian} = 0.678
\end{equation}
We have also shown that there possibly is another peak in the data
at $r \approx 0.78$, yet is was not significant.

\section{Comparison with the later study}

\begin{figure}
\resizebox{\hsize}{!}
{\includegraphics[angle=-90,width=0.9\columnwidth]{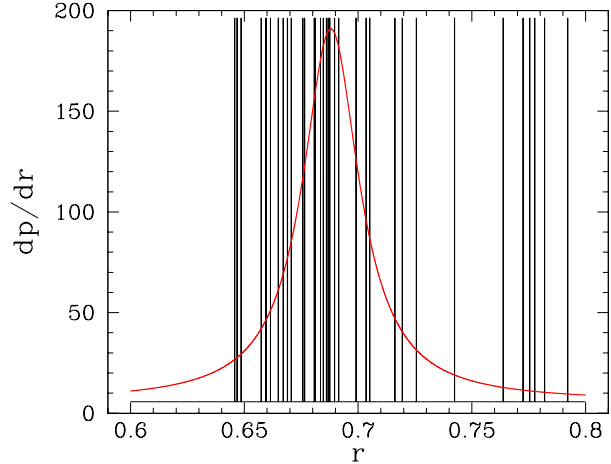}}
\caption{The frequency ratios used by \citet{2003A&A...404L..21A} are represented by the vertical lines. The solid line corresponds to the best fit single Lorentzian.}
\label{barcode}

\end{figure}
In a recent paper \citet{2005A&A...437..209B}
criticize the results obtained above.
There is a relation between 
the two kHz QPO frequencies, and therefore
it is equivalent to analyze the distribution of one or
of another frequency. The relation 
between the two kHz QPO frequencies is 
clearly visibly and could already be noted 
in   \citet{1997ApJ...481L..97V}. 
The statement would have been true if the
relation was exact, however this is just a correlation
and not a strict one to one correspondence.
One can verify that by checking the consistency of the distributions
of the lower and upper frequency with the hypothesis 
that they are distributed uniformly,
in an interval between the lowest the highest frequency in a given 
set. For the upper frequency we obtain the same result
as \citet{2005A&A...437..209B}, i.e. $P_{\rm upper}=0.017$,
while the analysis of the distribution of the 
lower frequency gives  $P_{\rm lower}=0.0012$.
In their conclusions
\citet{2005A&A...437..209B} write that 
the "distribution of QPO frequencies used by Abramowicz etal 2003
differs from constant can only be rejected at 2.4$\sigma$ level"
This statement is only true for the distribution of
the upper frequency and not for the lower. Moreover
 we never considered the distributions of individual 
 frequencies  in \citet{2003A&A...404L..21A} but their ratio. 
 The distribution of the frequency ratios differs
 from uniform at a level of $3.5\sigma$.
 Therefore the statement that "the results of 
 Abramowicz et al. 2003  are not statistically founded" is false.

\citet{2005A&A...437..209B} analyze the distribution of 
the upper kHz QPO in Sco X-1 using a larger data set. They found
that the distribution of this QPO is not compatible 
with a uniform level and the fit the distribution 
using three gaussians. It is interesting to note
that in all cases there  is a peak in the distributions they obtain 
corresponding to the 3/2 ratio of the frequencies assuming 
the linear relation between the two frequencies.
The distribution of the frequency ratios
can be transformed to the distribution 
of an individual frequency using the lines relation. 
Using the propagation of errors we estimated that 
there should be a peak in the distribution of upper QPO 
in Sco X-1 at the frequency 
\begin{equation}
\nu_{peak} = 940 \pm 25 {\rm Hz}.
\end{equation}
\citet{2005A&A...437..209B}
found a peak at $432.5\pm 1.5$\,Hz in complete agreement with
our results. 
It is extremely interesting to find that
the distributions of QPO frequencies in other sources
analyzed by \citet{2005A&A...437..209B} also 
show clustering and peaks at the location consistent with the value of
3/2, when the  linear relation between the two 
kHz QPOs is taken into account.

\citet{2005A&A...437..209B}
propose an explanation for the clustering 
of frequencies as a simple randon walk. In their model 
they start from 700Hz and let the frequency wander in frequency
in steps of $\pm 6$\,Hz. They show that this
model qualitatively explains clustering of frequencies.
This model is not defined in detail,
as it has not been clearly stated  for how many steps
the frequency is allowed to wander, or if there are any boundaries 
outside of which the frequency can not drift. 
Moreover in a real observation the starting point depends on 
a particular moment of the observation.
The distributions of the upper kHz QPO of \citet{2005A&A...437..209B}
is similar to the of  \citet{1997ApJ...481L..97V}, and the
Kolmogorov-Smirnov  test shows that the probability that they come from the same underlying distribution is 22\%.
One can ask a question how probable is it to obtain 
two such similar distributions in two 
realizations of a random walk. We have preformed 
a simulation and found that the probability 
of obtaining two such similar distributions varies between
$10^{-4}$ and $10^{-2}$ depending on the specific
prescription of the random walk.

\section{Summary and future observations}

We have shown that the analysis of the distribution
of the upper kHz QPO frequencies in Sco X-1 by 
\citet{2005A&A...437..209B} leads to the 
results in total agreement with the results
of \citet{2003A&A...404L..21A}. The peaks
in the distribution of upper kHz QPO frequency
found by  \citet{2005A&A...437..209B}
correspond exactly to the estimates based 
the earlier results of \citet{2003A&A...404L..21A}.
The deviation from a uniform 
distribution of the frequency ratios 
is at the $3.5\sigma$ level, and not $2.4\sigma$, which corresponds
to the distribution of the upper kHZ QPO
as claimed by \citet{2005A&A...437..209B}.
the distribution of the lower kHZ QPO differs
from uniform at a $3.2\sigma$ level. 
The difference comes from the
fact that the linear relation between the two frequencies
is not exact but has some spread.

The random walk model proposed by \citet{2005A&A...437..209B}
reproduces qualitatively the distribution 
of upper KHz QPO. The problem that it faces 
comes from the fact  that it is very difficult to
reproduce a similar distribution in two separate 
observations, while the set of data in \citet{1997ApJ...481L..97V}
is consistent with the set of \citet{2005A&A...437..209B}.
This model can be verified with a long term monitoring of
sources with kHz QPOs and a uniform sample of their behavior.
If the frequencies, and frequency ratios really do cluster
on some values such behavior should be reproducible in consecutive 
long observations. In the random walk model 
the peaks in the distributions of frequency ratios will wander from one observation to another.
Such study must be treated with caution 
as we have no information on the fundamental scale
of variability of kHz QPOs. 

Furthermore in my opinion there is no preferred 
way of analyzing the data. There is no reason a priori 
to say that the frequencies or frequency ratios, 
their sum or difference is in any way a better quantity to analyze
until we have a physical model to consider.
\citet{2003A&A...404L..21A} based the analysis 
on such a physical model in which the frequency ratio 
plays a fundamental role. The analysis 
by \citet{2005A&A...437..209B} is phenomenological 
and does not refer to any underlying 
physical model.

Finally there is the issue of the consistency of the 
data with the resonace theory \citep{2001A&A...374L..19A}.
The theory predicts a preference for the 2/3 ratio of the frequencies but 
is not saying anything yet about the shape of the
expected distribution around this value. 
Such shape should probably depend on the 
details of the interaction in the inner disk.
At this point we can only say that the value
of the ratio $r=2/3$ lies within the width of the assumed
shape of the distribution of frequency ratios in Sco~X-1.
This observation provides an additional hint
to consider the resnance model in more detail.

\acknowledgements

The author thanks for the hospitality and support 
from Nordita and the
KBN grant 2 P03D 001 25.

\newcommand{\aap}{Astronomy and Astropysics}

%\bibliography{qpo}

\end{document}